# Using Additional Indexes for Fast Full-Text Search of Phrases That Contain Frequently Used Words

A. B. Veretennikov

Ural Federal University
Institute of Mathematics and Computer Sciences (IMCS)
alexander@veretennikov.ru, AlexanderBorisovich@urfu.ru

**Abstract.** Searches for phrases and word sets in large text arrays by means of additional indexes are considered. Their use may reduce the query-processing time by an order of magnitude in comparison with standard inverted files.

**Keywords**: full-text search, search engines, inverted files, additional indexes.

## INTRODUCTION

Inverted files or their analogs are commonly used in searching for words or phrases [1–6]. An inverted file is a set of records of the form (ID, P), where ID is the document identifier and P is the position of a word (for example, the ordinal number of the word in a document). All records corresponding to a single word are stored in sequence so as to permit rapid reading during searches. Words in documents are encountered with different frequencies. The maximum time to respond to a search query, which is a very important parameter, is determined by the most frequently used words. Accordingly, we need to hasten the search for phrases that contain common words. To that end, we may create additional indexes.

In the present work, we consider the search for phrases or word sets in texts. The search query is a several words, and the search result is the list of documents with the indication of positions where the specified words may be found. We consider both precise search for phases and searches in which distance is taken into account. In the latter case, we look for documents where the target words are as close together as possible. That requires the storage of information in an index regarding each occurrence of each word in the documents.

This work is a continuation of [7], where we distinguished between three groups of words.

1. Stop words: *and*, *at*, *or*. These are very commonly used and may not be included in the index. For example, prepositions. In what follows, such words will be called stop words even if, in some form, we include information regarding these words in the index.

2. Frequently used words. These words are frequently encountered but convey meaning and should be included in the index.

3. All other words are classified as ordinary words.

In creating the index, we use a morphological analyzer. For each word form in the dictionary, the analyzer provides a list of numbers of the basic word forms. The number of the basic form is in the range from zero to *WordsCount* − 1, where *WordsCount* is the number of different basic forms (around 200 000 for Russian in the dictionary employed).

If the word does not appear in the analyzer's dictionary, we assume that its basic form is the same as the word.

When using the analyzer, we use three groups division approach not to words itself but to the basic forms of the words. In other words, there are three types of basic word forms in terms of the frequency with which they are encountered: stop basic forms, frequently used basic forms, and others.

The presence of several basic forms often complicates the software, and in some cases it makes sense to describe the algorithm without becoming involved in the technical details.

In the present work, we assume that any words may be encountered in the search query, regardless of their frequency of occurrence.

**NOTATION**

$*$, multiplication;
$<<$, bitwise left shift, used for integers with no sign;
$|$, bitwise *or*, used for integers with no sign;
$|X|$, modulus of the number $X$;
$<=$, less than or equal to, in the comparison of two numbers;
$!=$, not equal to;
*Record*.*Field,* identification of the field *Field* corresponding to the record *Record*. A record is a standard type of data, associated with several named fields.

**STRUCTURE OF THE INDEX**

In describing basic concepts, we will not specify the implementation of the index. However, insofar as technical details are described, we will assume the index structure described in [7].

For the basic form of the word, we define a stream as the list of records (ID, P) regarding its occurrences in the documents. The record (ID, P) is named as *posting*. Those records are stored sequentially in the index. The stream is described by a small structure, a descriptor, in which information regarding the location of the stream data in the index file is stored.

**SEARCH INDEXES FOR PHRASES CONSISTING OF STOP WORDS**

In the case of a query consisting entirely of stop words, we may use the additional stop word indexes described in [7–12]. Such an index stores all the information regarding the occurrences of any adjacent stop words *w*, *v*. However, the number of records may reach 3–30 millions in an index created for 30 GB of text, as shown in [7]. The problem is that if the search query contains $n$ stop words (3, 4, or more), we need to consider $O(n^2)$ streams for the search in this case.

Consequently, we need to create an index containing information regarding the appearance of all phrases consisting of stop words.

We introduce the parameters *MinLength* < *MaxLength*, and construct indexes for phrases of length $L$, where *MinLength* <= *L* <= *MaxLength*.

For example, if the text has 10 stop words arranged in sequence, we will have nine phrases with 2 words, eight phrases with 3 words, seven phrases with 4 words, and so on.

To create the index, we use a B-tree [13]. The key to the tree is the list of numbers of stop words (i.e. a phrase that consists of several stop words converted to a list of this stop word's numbers). The value of the key is a reference to an inverted index in which information regarding the occurrences of the corresponding phrase is stored. The order will be disregarded in the search. Therefore, the lists are sorted in increasing order.

In all, there are *MaxLength* − *MinLength* + 1 indexes.

Note that, for a query consisting entirely of stop words, only a precise search is considered. There will be no other words between the target words in the fragment of text that is found. This may be justified for the following reasons.

1. These words are often encountered. For practically every combination of such words, there will be an appearance in texts.

2. A significant proportion of such queries may relate to set phrases in which the composition and order of the words is fixed.

3. Such a query may be generated by copying a phrase from already existing text so as to search other documents that may include the phrase.

## ALGORITHM FOR INDEX CREATION

We now describe the algorithm when using a morphological analyzer.

We assume that we have a stop list consisting of all the stop basic forms.

We create a queue *Queue,* whose elements are records *QueueItem* with three fields (ID, P, *Forms, Index, Next*). Here *Forms* is a list of the numbers of the words' basic forms; *Index* is an auxiliary variable; and *Next* is the next element of the queue.

When we read files for each occurrence (ID, P) of a word *w*, we form the list *Forms* of the basic word forms of *w* that appear in the stop list.

If the *Forms* field is not empty, we add the record (ID, P, *Forms*) to the end of *Queue*. Then, if the length of *Queue* exceeds *MaxLength,* we remove the first element. We call up the function *Process*(Begin of Queue*,* 1).

If *Forms* is empty, then, as long as *Queue* is not empty, we call up the function *Process*(Begin of Queue, 1) and remove the first element in the queue.

Then we write the function *Process*(*Item, L*)*,* where *Item* is the record *QueueItem* and *L* is the length of the fragment.

In cycling through the list *Item.Forms*, the current index in the list (cycle counter) is stored in *Item.Index*. In each iteration of the cycle, the following steps are performed.

1. *Process*(*Item.Next, L+*1).

2. If *MinLength* <= *L* <= *MaxLength,* the next steps are performed. Otherwise, the processing of the function ends.

3. Formation of the list *WordIDs* of forms that include, in sequence, *Forms*[*Index*], beginning at the start of the *Queue*, and processing of the *L* elements in the *Queue* (*Current* = Begin of the Queue; *L* is the number of repetitions: *Current.Forms*[*Current.Index*] is added to *WordIDs, Current = Current.Next*).

4. Replacement of all the numbers of basic word forms in *WordIDs* by the corresponding numbers in the stop list.

5. Sorting of *WordIDs* in ascending order; coding by the Huffman algorithm to reduce the size.

6. Storing of record (ID, P) in the index, using *WordIDs* as the key.

## OPTIMIZATION OF SEARCH-QUERY PROCESSING USING EXPANDED INDEXES

The expanded index (*w, v*) is a list of occurrences of the word *w*, when word *v* is present in the text at a distance less than *ProcessingDistance* from *w*, as described in [7] (*w* – frequently used, *v* – frequently used or ordinary). The value of the parameter *ProcessingDistance* depends on the frequency of occurrence of the word *w* in texts. We assume that, if the distance between the words is less than *ProcessingDistance*, the words are related in meaning, but otherwise they are not.

Note that if an expanded index exists for both words *w* and *v* — that is, if indexes (*w*, *v*) and (*v*, *w*) exist — it is sufficient to create one of them — say, (*w*, *v*) — and to save the distance between *w* and *v* in the posting. By that means, the size of the index may be reduced.

## EXPANSION OF INFORMATION STORAGE REGARDING STOP WORDS IN THE INDEX

In addition to the stop word index and the index for frequently used words, we use a basic index.

All the occurrences of frequently used words and ordinary words are stored in the basic index.

In the basic index, for each word, we may store not only its occurrence but also information regarding near stop words, as noted in [7]. This approach may be expanded to storage of information regarding stop words that are no further than *MaxDistance* from the word being considered. Experiments show that the associated increase in size of the index is acceptable.

For example, suppose that we select *MaxDistance = 5*.

Then we will store all occurrences of the frequently used words and ordinary words in the basic index. For each occurrence, we will also store information regarding near stop words.

If the word is frequently encountered, a separate stream may be used to store information regarding the stop words. In this case, then, we may have up to three streams for the storage of information regarding occurrences of the word.

1. Storage of the ID of the document + the first occurrence in this document + the number of occurrences in the document.
2. Other occurrences.
3. Information regarding the near stop words for each occurrence.

With this approach, there is no need to read the information regarding the stop words if it is irrelevant to the query at hand.

The first two streams were employed in [7]. On that basis, the same index may be used for all searches, regardless of whether the distance is of interest. Searches without consideration of the distance are significantly faster, since only the first stream is employed, and the number of operations that involve the records is an order of magnitude less. For rarely used words, all the data may be stored in a single stream. That reduces the number of input/output operations required in creating the index.

## ANSWERING QUERIES

We now consider the fulfillment of various types of queries. In the examples, we only mention those streams containing lists of the occurrences of words relevant to the search.

**Type 1.** All the words in the query are stop words.

We use the stop word index. We form the key as a sorted list of the IDs of stop words (its basic forms) and, on that basis, we extract all the occurrences from the index.

Example 1: not only that but.

Example 2: which it would be if.

**Type 2.** All the words are frequently used.

Here we consider a query in which there are no stop words and all the words are frequently used.

First, we consider the case where all the words in the query are frequently used. In other words, we have an expanded index for each word.

We select the word $w_i$ in the query that is encountered least often in texts. Then, to search for the query, it is sufficient to consider $n - 1$ expanded indexes. For each word $w_k$, $k \mathrel{!}= i$, we consider the index $(w_k, w_i)$.

Example: rivers define boundaries.

The word <boundaries> is least often encountered in texts.

1. (river, boundary): search in the expanded index.
2. (define, boundary): search in the expanded index.
3. boundary, define, river: search in the basic index, first occurrence in document.

Here and in what follows, we first take account of the distance (steps 1 and 2 in the example). Then, if no result is obtained, we disregard the distance (step 3). For the search disregarding the distance, we need an index in which only the first occurrence of the word in the document is retained. That reduces the number of occurrence records by an order of magnitude, as noted in [7].

**Type 3.** Not all of the words are frequently used and there are no stop words.

We now consider the case where at least one word in the phase is neither a stop word nor frequently used. Suppose that the query includes $m$ such words. We select the word $w_i$ that is encountered least often in texts. We call it the *basic word* in the query. Then, to search for the phrase, it is sufficient to consider $n - m$ expanded indexes. For each word $w_k$, $k \mathrel{!}= i$, and an expanded index $(w_k, w_i)$ exists we consider the index $(w_k, w_i)$. In the case of words for which no expanded index exists, we use an ordinary index.

In extracting records from the basic index, information regarding near stop words is passed over, if possible (that is, if it is stored in a separate stream).

Example: fragrant red rose.

Here the word <fragrant> does not appear in the list of frequently used words. Two basic forms exist for <rose>: rise, rose. The basic query word is <fragrant>.

1. (red, fragrant): search in the expanded index.
2. (rise, fragrant), (rose, fragrant): search in the expanded index.
3. fragrant, red, rise, rose: search in the basic index, first occurrence in document.

**Type 4.** All forms of the word appear in the phrase.

We proceed as in the previous case. But for the selected basic query word $w_i$, we consider all its occurrences and process information regarding near stop words in order to take account of stop words appearing in the phase. In that case, the basic word in the query may also be a frequently used word. If so, we take all its occurrences from the basic index.

Example: reports about Gallic war.

The stop words in this query are <about> and <war>. <Report> – frequently used, <gallic> – ordinary.

1. gallic: search for all occurrences with analysis of information regarding stop words.
2. (gallic, reports): search in the expanded index.

The question that arises here is whether to search without taking account of the distance, since such a search may not make sense for the stop words. Where necessary, this search may be undertaken.

In the following example, frequently used words and stop words appear.

Example: all necessary things for the walk.

The stop words in this query are <all>, <for>, and <the>.

1. necessary: search for all occurrences, with analysis of information regarding stop words.
2. (necessary, thing).
3. (necessary, walk).

## PROCESSING QUERIES

For each word in the query, the morphological analyzer gives several basic word forms. Thus, for a query consisting of $n$ words, there are $n$ lists of basic forms. If one of the lists includes basic forms that differ in the frequency type — say, a stop form and a frequently used form — the query must be divided into two parts, each of which is processed separately. Otherwise the search algorithm is significantly complicated.

To explain in more detail, suppose that we have a list *Query*, each element of which is a list of basic word forms. We consider each element of *Query*. Suppose that $i$ is the number of the element considered. If it contains $m$ different types of basic forms in terms of the frequency with which they occur, we create $m$ copies of the search query, in each of which element $i$ contains only a single type of basic form (stop forms, frequently used or ordinal). For each new query, we proceed analogously, processing element ($i + 1$) and those that follow. Then the results for each query are obtained and combined.

**EXPERIMENTS, RESOURCES, AND INITIAL DATA**

All the experiments are conducted using a collection of tests of magnitude 45 GB. These documents take the form of simple text. In style, they correspond mainly to fiction and magazines. In all, we have around 130 000 documents.

For the search experiments, we use the following resources.

CPU: Intel(R) Core(TM) i7 CPU 920 @ 2.67GHz.
HDD: Seagate Barracuda 7200.11, 7200 RPM, cache 32 MB, 2 GB, ST32000641AS.
RAM: 24 GB.
OS: Microsoft Windows 2008 R2 Enterprise.

The parameters of the test index are as follows: *MinLength* = 2, *MaxLength* = 5, *MaxDistance* = 5–7, depending on the frequency with which the word is encountered.

We identify 700 stop words that are most commonly encountered. Usually, if they are not included in an index, the list of stop words will be as short as possible. Since we do include them in an index, this list may be relatively long.

We have 2100 frequently used words.

The given parameters permit searches for phrases with up to five words. In the search for longer phrases, containing stop words or frequently used words, the phrase may be divided into parts. Each part is then processed separately, and the results are combined.

**STRUCTURE OF SEARCH EXPERIMENTS**

The experimental procedure is as follows.
1. Selection of a random document in the index.
2. Selection of search phrases as follows.
2.1. Selection of a sequence of words.
2.2. Selection of a sequence of words, with the omission of every other word. For example, consider a document «Gaul, taken as a whole, is divided into three parts…»: we select queries «Gaul taken as» at 2.1 and «Gaul as whole» at 2.2, then «taken as a» at 2.1 and «taken a is» at 2.2, and so on.
3. Search for each selected set of words. In the search, all the records corresponding to the given word are read. Thus, even if the required set of words is found, reading continues to the end.

Sets of three, four, or five words are selected.

If one of the query words has a stop basic form, the search is confined to sequential words.

The basic parameter measured is the number of postings regarding word occurrences that are read in processing a single search query. The search rate mainly depends on this parameter. After a large number of searches, the mean and maximum values of this parameter are determined. The time to process the query is also determined.

The benefits of this approach are as follows.

1. We verify that the index is correctly constructed and performs as required. Since phrases are selected from an already-indexed document, they should be precisely found. We also verify that the search results include a record corresponding to the document used in selecting the query.

2. The phrases found are relatively diverse and include a large number of different words. Many of the phrases include stop words and frequently encountered words.

All the queries are processed sequentially in a single program thread. In other words, a single processor core is used.

**SIZE OF THE INDEXES**

| Index | Size |
| --- | --- |
| Index used in searches for stop-word phrases | 80 GB |
| Expanded index used in searches for phases including frequently encountered words | 79 GB |
| Basic index | 67 GB |
| Total (all indexes, metadata, compressed document texts) | 259 GB |

**SEARCH SPEED**

In all, 45 000 queries are processed, in 1 h 38 min. All possible types of words are encountered in the queries.

| Characteristic | Mean value | Maximum value |
| --- | --- | --- |
| Time to process query | 0.13 s | 1.31 s |
| Number of queries processed | 274 000 | 6 million |

An ordinary index is also constructed for the same data set by means of Sphinx 2.0.6 software [14]. The size of the index is 18.7 GB. The same queries are processed.

| Characteristic | Mean value | Maximum value |
|---|---|---|
| Time to process query | 1.01 s | 17.82 s |
| Number of queries processed | 112 million | 505 million |

**CONCLUSIONS**

We have proposed methods of organizing additional indexes for stop words and frequently encountered words, with a goal to increasing search speeds.

In experiments, the maximum search time for various phrases that include stop words and frequently encountered words is an order of magnitude less when using additional indexes than when using ordinary indexes.

In experiments regarding the creation of an index that incorporates additional indexes of frequently used words, we find that the size of the additional indexes, with optimal parameters, is around 250 GB when indexing texts of magnitude 45 GB. The increase in index size may be acceptable if the search speed is significantly increased.